\documentclass[12pt,preprint]{aastex}

\begin{document}

\title{Contamination by Surface Effects of Time-distance Helioseismic Inversions for Sound Speed Beneath Sunspots}

\author{Couvidat, S. \& Rajaguru, S. P.}

\email{couvidat@stanford.edu}

\affil{W.W. Hansen Experimental Physics Laboratory, Stanford 
University, Stanford, CA 94305-4085}

\received{updated 2/8/2007}

\shorttitle{Contamination by Surface Effects of Time-distance Helioseismic Inversions}
\shortauthors{Couvidat}

\begin{abstract}
Using Doppler  velocity data from  the SOI/MDI instrument  onboard the
SoHO spacecraft, we do time-distance helioseismic inversions for 
sound-speed perturbations beneath 16 sunspots observed in high-resolution  
mode. We clearly detect ring-like regions of enhanced sound speed beneath 
most sunspot penumbrae, extending from near the surface to depths of about
$3.5$ Mm. Due to their location and dependence on frequency bands of p-modes
used, we believe these rings to be artifacts produced by a surface signal
probably associated with the sunspot magnetic field. 
\end{abstract}


\section{INTRODUCTION}

Time-distance helioseismology (Duvall et  \mbox{al.} 1993) has a track
record of  significant findings in solar physics:  from the subsurface
structure  and flow  fields of  sunspots (\mbox{e.g.} Duvall  et  \mbox{al.} 1996;
Kosovichev et \mbox{al.} 2000; Zhao et \mbox{al.} 2001) to the mapping
of large-scale  subsurface flows (Haber et \mbox{al.}  2002), and even
to  the in-progress  far side  imaging (Zhao,  private communication).
These results are  based on the assumption that the acoustic wave travel
times are only marginally affected by the presence  of  magnetic
fields. Recent work in sunspot seismology, based largely on helioseismic
holography techniques, however, point to a dominant influence of surface
magnetic field on two related phenomena: 
the so-called showerglass effect (Lindsey \& Braun 2005), and
the  effect  of  an  inclined  magnetic field  (\mbox{e.g.}  Cally  et
\mbox{al.} 2003;  Schunker et \mbox{al.}  2005; and Schunker  \& Cally
2006). A clear understanding of the origin of these surface effects
is still lacking, and recent results 
point to contributions from several distinct
sources: while Schunker et \mbox{al.}  (2005), and associated modelling
work (Schunker  \& Cally 2006), emphasize the importance of wave propagation
in inclined magnetic field to the azimuthally varying local
control correlation phases (Braun \& Lindsey 2000) over penumbral regions,
Rajaguru et \mbox{al.}  (2007) show that such wave propagation effects
could manifest through altered height ranges of spectral line
formation that affects Doppler shift measurements within sunspots. 
While such surface effects have been confirmed present in time-distance 
measurements too (Zhao \&  Kosovichev 2006), these latter authors 
conclude that they do not change appreciably the previously obtained
deeper extending two-region (cold-hot) structure of sound-speed perturbations
below a sunspot and thus stress the presence of real subphotospheric structures. 
Here we analyze  16 active  regions of  different sizes  and magnetic
field fluxes: we compute travel-time perturbation maps for 11 travel distances 
and invert them for 3-dimensional sound-speed perturbations beneath each active region. 
We show the presence of ring-like structures of significant positive sound speed changes,
down to a depth of about  $3.5$ Mm, that seem 
likely to be artifacts due to our ignoring the surface magnetic effects in the
inversion procedures. We perform experiments to check the origin of these surface
effects: the amplitude effect [interaction of phase-speed filter with
localised reduced amplitudes of oscillations within sunspots, Rajaguru et \mbox{al.} (2006)]
is shown not to account for it, the location and frequency dependence of the ring-like
sound speed excesses are found to be related to similar dependences of travel times observed recently
by Braun \& Birch (2006), and the ring-like structures are shown to be present only on the outgoing travel-time maps.

In section 2 we describe the methodology applied here to obtain travel-time
maps and to invert for the sound-speed perturbation. In section 3 we present
our results. We conclude in section 4.

\section{TIME-DISTANCE ANALYSIS}
\subsection{Datasets}

We use high-resolution dopplergrams derived from filtergrams of circular
polarization component of Ni {\sc i} ($\lambda$=6768 \AA) observed by the
Solar Oscillations Investigation/Michelson  Doppler Imager (SOI/MDI) onboard
SoHO (Solar and Heliospheric Observatory) (Scherrer et  \mbox{al.} 1995).
These  dopplergrams are mapped  using a Postel  projection and
are  tracked  at  the  Snodgrass  differential rotation  rate corresponding
to central latitude of the MDI hi-res field of view (mapping  and  tracking
are done using the ``fastrack'' program in the Stanford MDI/SOI pipeline software).
Dopplergrams  that  are  contiguous in  time  are  stacked
together   to  produce  3D   datacubes  $\phi(\mathbf{r},t)$   of  the
line-of-sight (l.o.s) Doppler   velocity  at   the   solar  surface,   where
$\mathbf{r}=(x,y)$  is the  horizontal  vector, and  $t$  is the  time
coordinate.     The    dimensions     of    these     datacubes    are
$256\times256\times512$ nodes, with a spatial resolution $dx=dy=0.826$
Mm, and  a temporal  resolution $dt=1$ minute.  The same  procedure is
applied  to produce  datacubes of  the l.o.s magnetic
field  $\mathbf{B}(\mathbf{r},t)$,  and  of  the  continuum  intensity
$I_0(\mathbf{r},t)$. The  continuum intensity is the  intensity in the
solar continuum as measured by MDI at the tuning position number zero,
corresponding to two symmetrical measurements 150 m\AA $\,$ away from the
Ni {\sc i} line center.  The Doppler velocity  datacubes are detrended
by a  linear fit of the  long-term temporal trend.  When a dopplergram
has  no data or  is corrupted,  we replace  it by  the average  of its
neighbors.

We have selected datacubes containing sunspots  of various  sizes  and magnetic  field
strengths. Unfortunately, due to a saturation effect in the MDI magnetogram
measurements in large sunspots, the absolute value of magnetic field strength
in the umbrae of such spots is smaller than the actual values present in them.
This phenomenon was explained by Liu, Norton, \& Scherrer (2007) as being due to  a
limitation in the onboard processing of dopplergrams and magnetograms.
It limits our access to very large sunspots.

\begin{figure}
\plotone{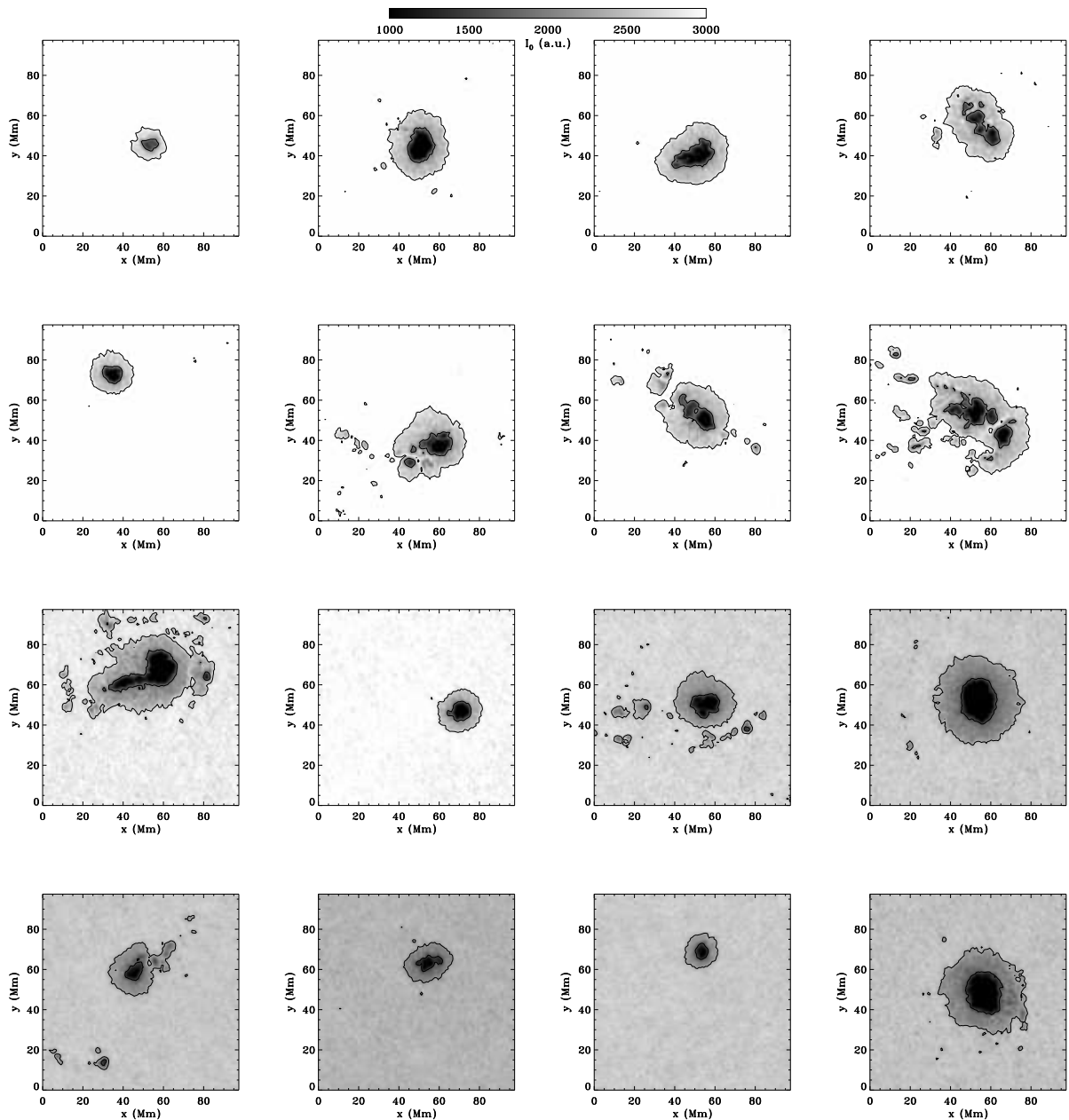}
\caption{MDI continuum intensity of the 16 solar active regions studied in this paper, in the order listed in Table \ref{table.spot} (from the upper left to the lower right panel). The boundaries of the umbra and penumbra are overplotted: they correspond to $57\%$ and $85\%$ of the continuum intensity, respectively. The greyscale is the same for every panel.\label{f1}}
\end{figure}

For the purpose of  this paper, a  sunspot is  defined as a  group of
contiguous pixels with a l.o.s magnetic field strength larger than 150
Gauss and that exhibits both an umbra and penumbra. The area of a spot
is only the  number of pixels within this spot  multiplied by the area
of a pixel $dx \times  dy=0.682$ Mm$^2$. The l.o.s magnetic field flux
is the value of  $\mathbf{B}(\mathbf{r},0)$ at each pixel muliplied by
the area  of a pixel. Table \ref{table.spot} summarizes the properties
of the  active regions studied in this paper, while  Figure \ref{f1}
shows maps of their continuum intensity $I_0$.

\subsection{Travel-Time Estimation}

The time-distance formalism  was introduced  by Duvall  et \mbox{al.}
(1993). It is based on computation of  cross-covariances between
the solar oscillation signals at two locations on the solar surface (a
source  at $\mathbf{r_1}$  and a receiver at $\mathbf{r_2}$).  Due to  the
stochastic excitation of acoustic waves by the solar convection zone and
due to the oscillation signal at any location being a superposition of 
a large number of waves of different travel distances (i.e., of different
horizontal phase velocities $v_{ph}=\omega/k$, where $k$ is the horizontal
wavenumber and $\omega$ is the temporal angular frequency), the cross-covariances 
are very noisy and need to be phase-speed filtered and averaged
(Duvall  et \mbox{al.} 1997). The Doppler velocity data cube $\phi(\mathbf{r},t)$ is phase-speed
filtered in  the  Fourier  domain, using  a Gaussian filter $F(k,\omega;\Delta)$  
for each travel distance $\Delta$. The standard scheme of averaging is to 
average the point-to-point cross-covariances over an annulus of radius 
$\Delta=|\mathbf{r_2}-\mathbf{r_1}|$ centered on the source. Such point-to-annulus
cross-covariances are  computed for several distances  $\Delta$ (55 in
this paper),  and then averaged by  groups of 5 distances to further
increase  their signal-to-noise  ratio. A detailed explanation of all these 
steps in the analysis process can be found in, \mbox{e.g.}, 
Couvidat, Birch, \& Kosovichev (2006), along with a table
of distances $\Delta$  and phase-speed filter characteristics used
(see Table 1 of previous reference).

The  point-to-annulus  cross-covariances   are  fitted  by  two  Gabor
wavelets (Kosovichev \& Duvall 1997): one for the positive times, one
for the negative times. We use the ingoing (i) and outgoing (o) phase travel times  
$\tau_{\mathrm{i/o}}(\mathbf{r},\Delta)$ returned by the fits. The average of
these  two travel  times, $\tau_{\mathrm{mean}}(\mathbf{r},\Delta)$, is,
in   first   approximation,   sensitive   only   to   the   sound-speed
$c(\mathbf{r},z)$ in the  region of the Sun traversed  by the wavepacket
(where  $z$ is  the  vertical coordinate). Using  a solar  model as  a
reference  we can  relate the  mean travel-time  perturbations $\delta
\tau_{\mathrm{mean}}(\mathbf{r},\Delta)$  to the sound-speed
perturbations $\delta c(\mathbf{r},z)$ through an integral relation:

\begin{equation}
\mathrm{\delta} \mathrm{\tau_{\mathrm{mean}}}(\mathbf{r},\mathrm{\Delta}) = \int\!\!\!\!\int_S d\mathbf{r'} \int_{-d}^0 dz \, K(\mathbf{r}-\mathbf{r'},z;\mathrm{\Delta}) \, \frac{\mathrm{\delta} c^2}{c^2}(\mathbf{r'},z) \label{eq1}
\end{equation}

where  $S$ is  the  area of  the region,  and  $d$ is  its depth.  The
sensitivity kernel for  the relative squared sound-speed perturbations
is  given  by $K$.  Here  we  use  Born-approximation kernels kindly 
provided  by \mbox{A. C.}  Birch (Birch, Kosovichev, \&  Duvall 2004). These
kernels  take  into  account  the  finite wavelength  effects  of  the
wavefield. Equation 1 is only  approximate because effects on the mean
travel times other  than the  sound-speed perturbation  are completely
ignored. For instance, Bruggen \& Spruit (2000) describe the impact of
changes in the upper boundary  condition in sunspots due to the Wilson
depression; Woodard (1997) and  Gizon \& Birch (2002) demonstrate that
increased  wave damping  in  sunpots can  introduce  shifts in  travel
times;  finally,  the significant  magnetic  field effects,  already  mentioned  in
Introduction, also  contribute to $\delta \tau_{\mathrm{mean}}(\mathbf{r},\Delta)$.

\subsection{Inversion of Travel-Time Maps}

Based on Equation \ref{eq1} the mean travel-time  perturbation maps (11 of them  per datacube) are
inverted  for  sound-speed  perturbation   using  a  regularized
least-squares scheme  based on a modified Multi-Channel Deconvolution
algorithm (MCD; see Couvidat et \mbox{al.} 2005 for details on the modified algorithm, and
Jacobsen et \mbox{al.} 1999 for the basic algorithm). The
modified   algorithm    includes   both   horizontal    and   vertical
regularizations. We use the weighted norm of, and that of the first 
derivatives of, the solution $\mathrm{\delta} c^2/c^2(\mathbf{r},z)$ for
vertical and horizontal regularization, respectively.  Following 
Couvidat et \mbox{al.} (2005) the covariance matrix of
the  noise  in the  travel-time  maps  is  included in  the  inversion
procedure, and  is obtained by  40 realizations of  noise travel-time
maps based  on the noise  model by Gizon  \& Birch (2004).   Both the
horizontal and vertical regularization parameters are constant for the
inversion  of  travel-time   maps  of  all  our  datasets.  These
parameters are chosen  empirically. We invert for 13  layers in depth,
with the center of the deepest layer located $28.5$ Mm below the solar
surface.

\section{RESULTS AND DISCUSSION}

\begin{figure}
\plotone{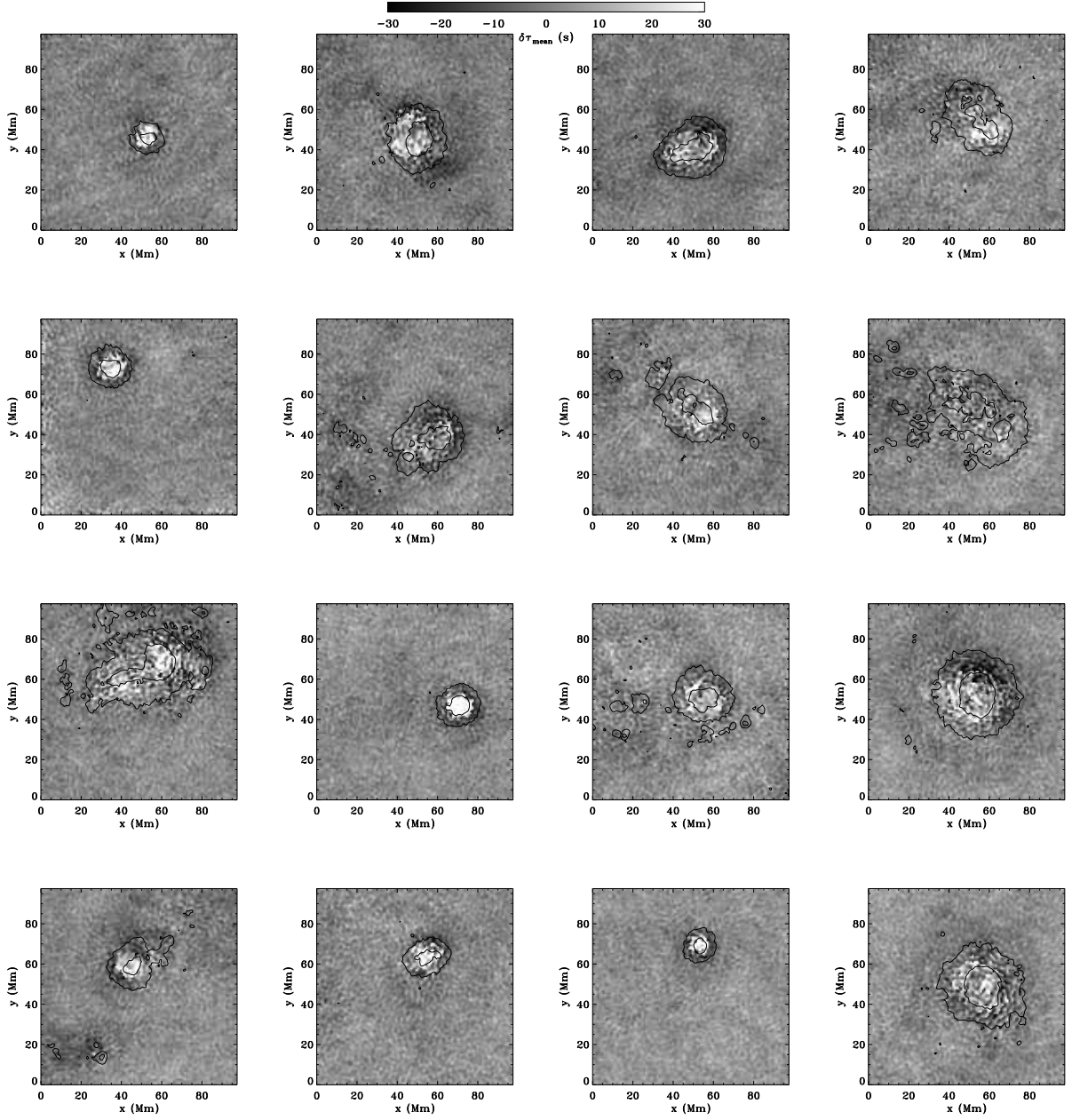}
\caption{Mean travel-time perturbation maps $\delta \tau_{\mathrm{mean}}(\mathbf{r},\Delta)$ for the distance $\Delta=11.6$ Mm and for the 16 sunspots studied. The greyscale is the same for every panel and was truncated to emphasize the dark rings. The boundaries of the umbra and penumbra, as defined on Figure \ref{f1}, are overplotted.\label{f2}}
\end{figure}

\begin{figure}
\epsscale{0.7}
\plotone{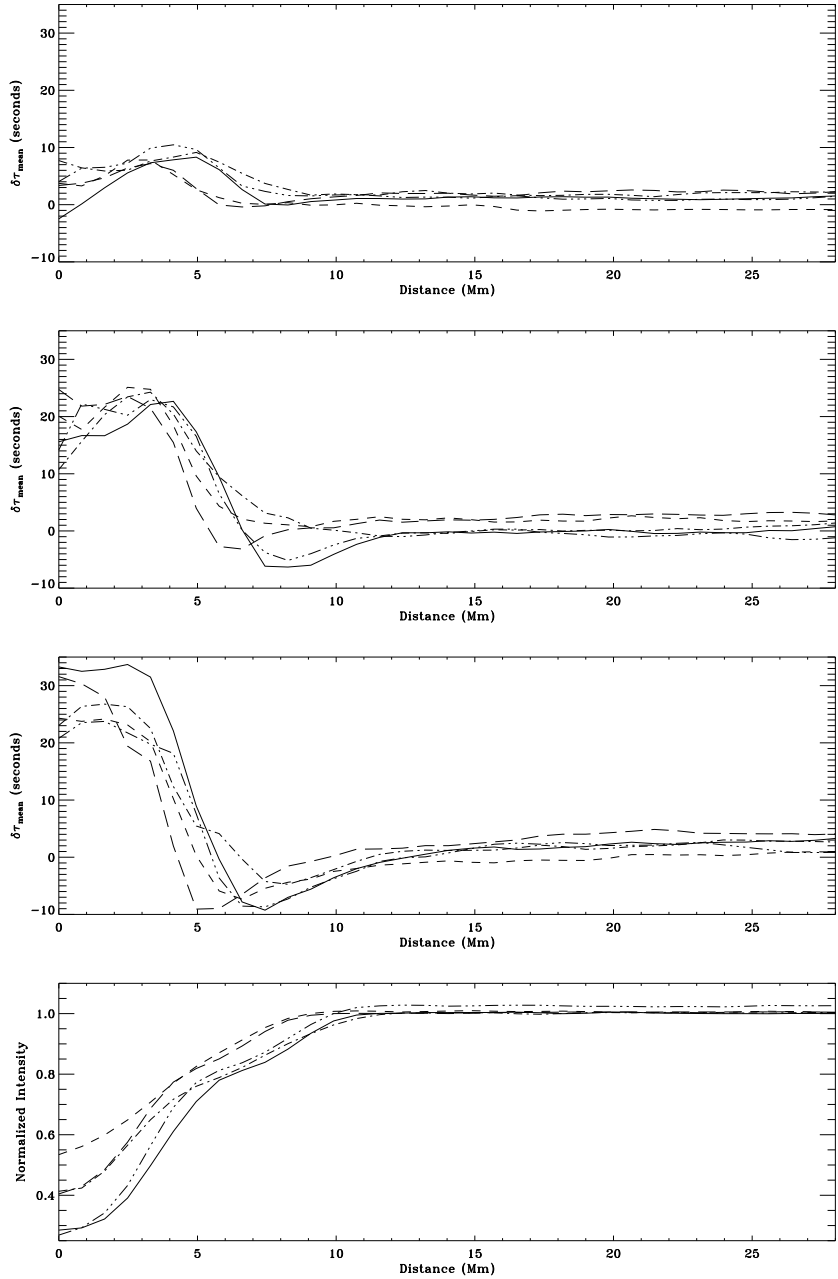}
\caption{Azimuthal average of $\delta \tau_{\mathrm{mean}}(\mathbf{r},\Delta)$ for $\Delta=6.2$ Mm  (upper panel), $\Delta=8.7$ Mm (second panel from the top), and $\Delta=11.6$ Mm (third panel) and for 5 sunspots: NOAA 9493 (solid line), 8073 (short-dashed line), 0615 (dot-dashed line), 8402 (dot-dot-dot-dashed line), and 0689 (long-dashed line). The lower panel shows an azimuthal average of the normalized continuum intensity.}
\label{f9}
\end{figure}

Figure \ref{f2}  shows the mean travel-time perturbation  maps  for
a travel distance $\Delta=11.6$ Mm over the 16 active regions. The main feature
visible  on most  maps is  a central  white disk  of  increased travel
times. Surrounding this disk, a dark ring corresponding to a decreased
mean travel time is clearly  present. Umbra and penumbra boundaries overplotted
on Figure \ref{f2} show that this dark ring is mostly located within the penumbral region of  the spot.
Such  dark  rings actually  appear in travel-time maps for 2  of the  11
distances  $\Delta$ studied  here: $\Delta=8.7$  Mm  and $\Delta=11.6$
Mm.        An        azimuthal        average        of        $\delta
\tau_{\mathrm{mean}}(\mathbf{r},\Delta)$ around the sunspot center for
5 circular sunspots shows the average amplitude of the rings and their
extent  (see Figure  \ref{f9}).  The rings do  not  seem to  be present  at
$\Delta=6.2$  Mm; some  spots exhibit  them but  most do not.   

\begin{figure}
\epsscale{1.0}
\plotone{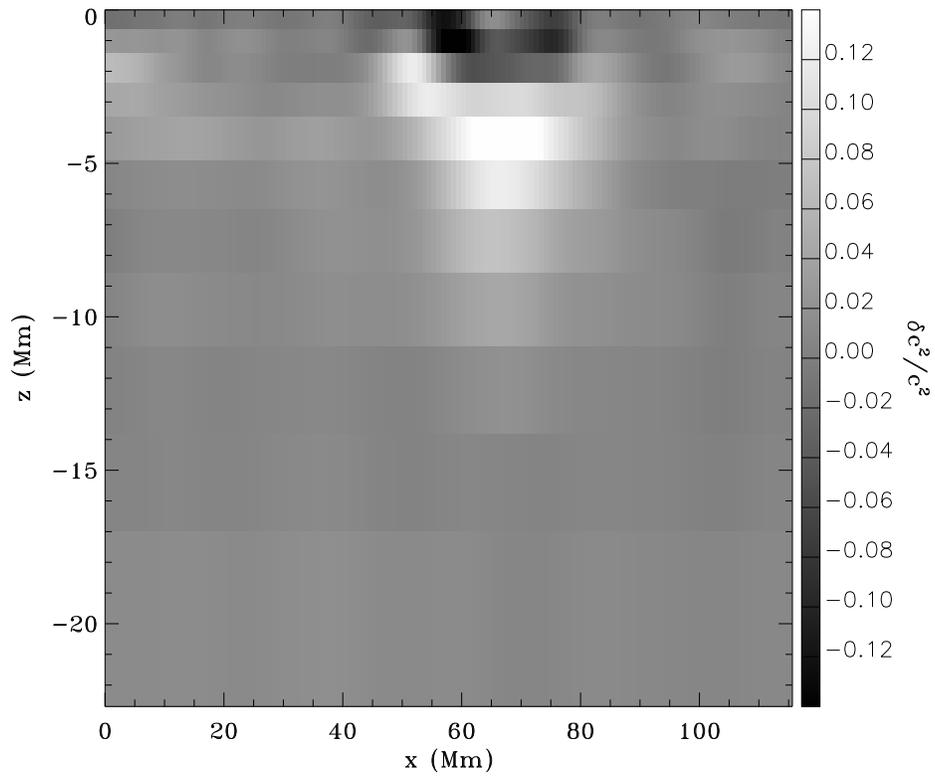}
\caption{Vertical cut in the inverted sound-speed perturbation for the active region NOAA 8397.\label{f6}}
\end{figure}

\begin{figure}
\plotone{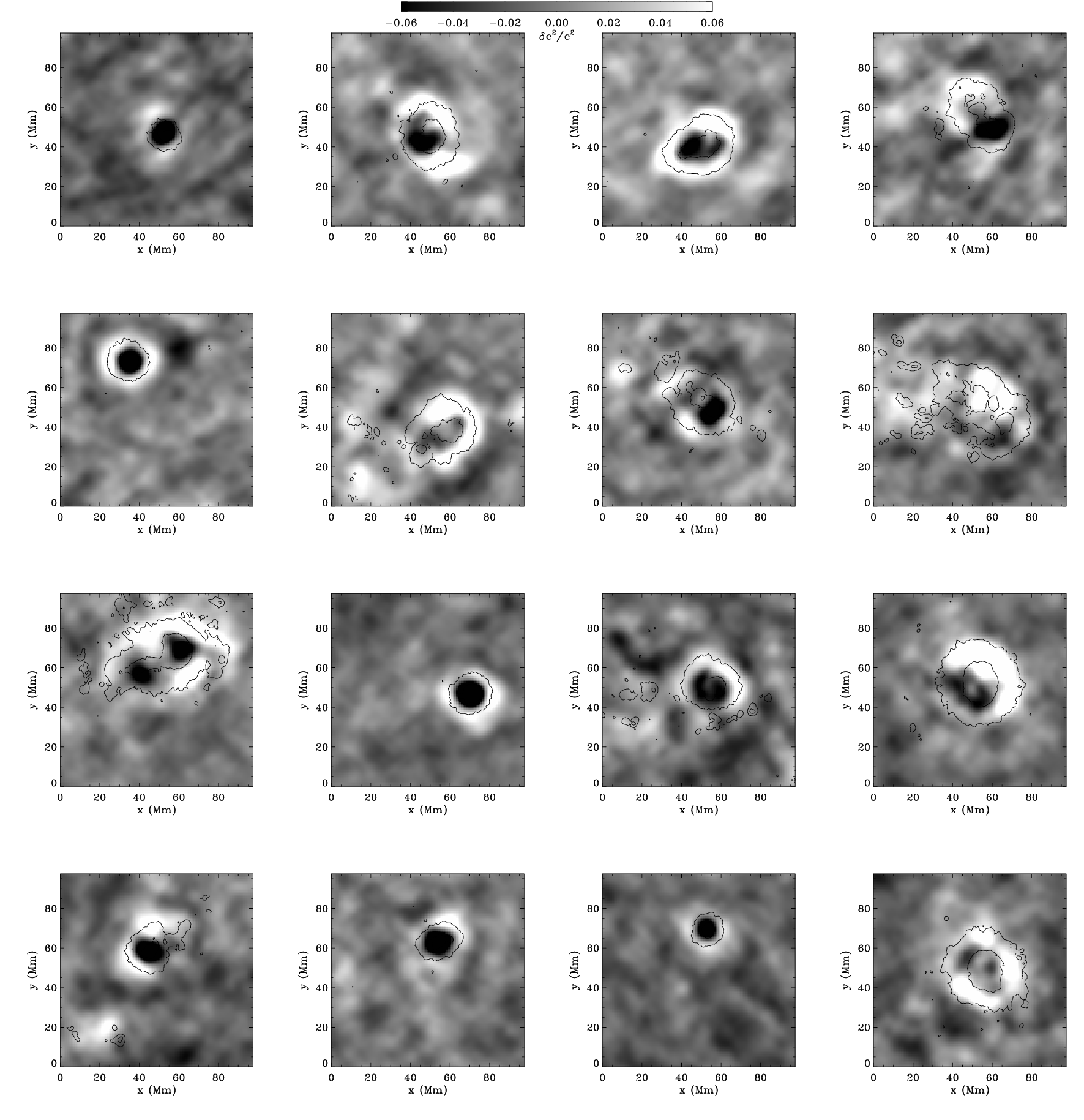}
\caption{Inversion results $\delta c^2/c^2$: we show a horizontal cut at a depth of $z=-2.38$ to $z=-1.42$ Mm. The greyscale is the same for every panel and was truncated to emphasize the bright rings. The boundaries of the umbra and penumbra, as defined on Figure \ref{f1}, are overplotted.\label{f5}}
\end{figure}

\begin{figure}
\plotone{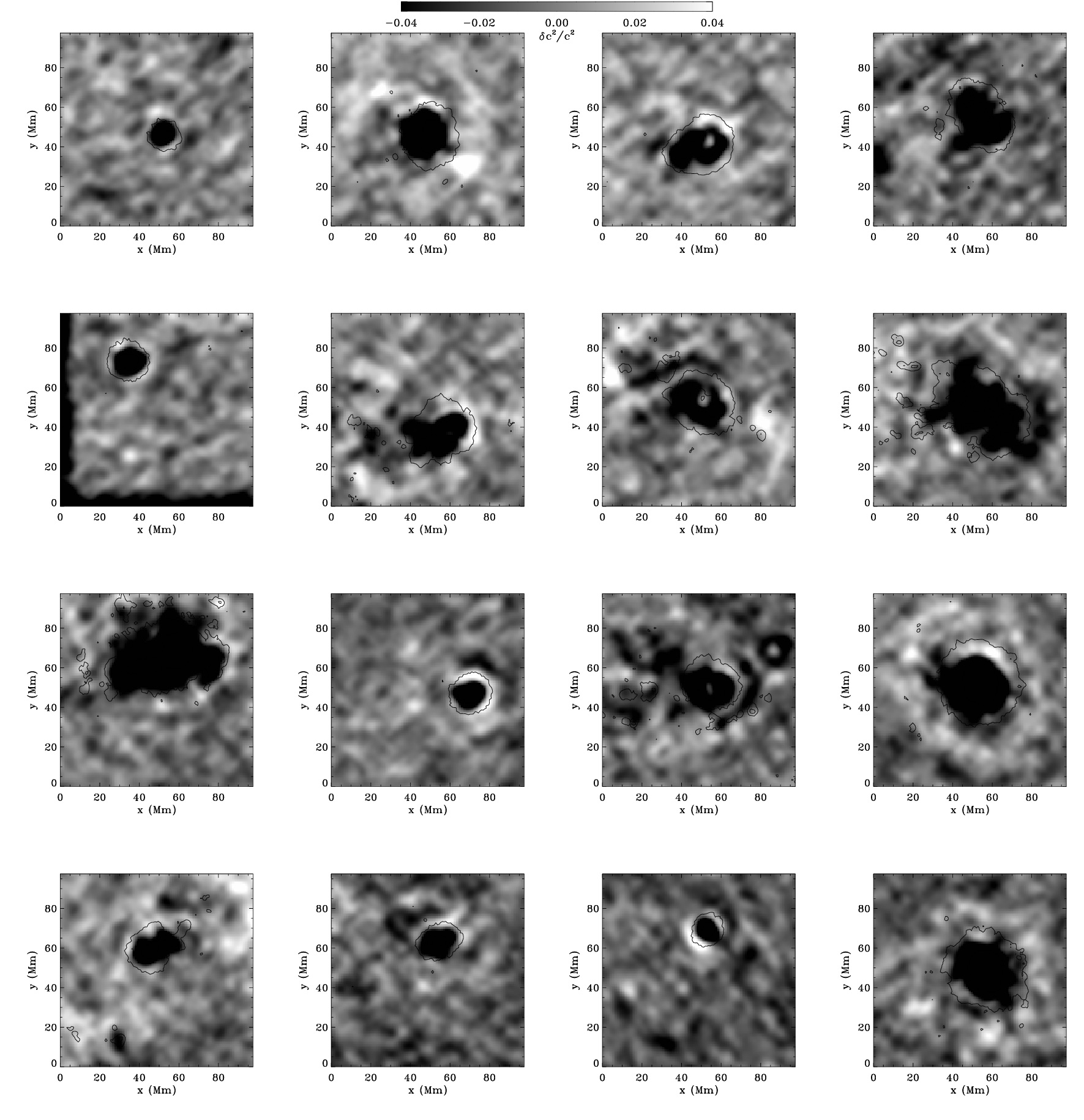}
\caption{Inversion results $\delta c^2/c^2$: we show a horizontal cut at a depth of $z=-1.42$ to $z=-0.62$ Mm. The greyscale is the same for every panel and was truncated to emphasize the bright rings. The boundaries of the umbra and penumbra, as defined on Figure \ref{f1}, are overplotted.\label{f8}}
\end{figure}

When travel-time maps  are inverted for sound-speed perturbations,
they  produce the  well known  two-region structure  below  the active
region: first the sound speed decreases just below the photosphere and
then increases  at a depth of  about 3-4 Mm, depending  on the sunspot
size  and magnetic  flux.  These two  regions  can be  seen in  Figure
\ref{f6}:  this is  a vertical  cut in  the inversion  result  for the
active region NOAA 8397.   The dark ring surrounding the corresponding
sunspot on the  travel-time maps of Figure \ref{f2}  appears on Figure
\ref{f6} as  two bright  arms extending from  the region  of increased
sound-speed toward the surface. These arms are especially well visible
at a  depth of  about $2$  Mm. Their amplitude  decreases when  we get
closer to the solar surface. The rings/arms are not visible anymore at
the  surface, in agreement  with the  travel-time map  at $\Delta=6.2$
Mm.   Such   ring-like/arm-like  structures   are   present  on   most
sunspots.  Figure  \ref{f5} shows  horizontal  cuts  in the  inversion
results for the layer $z=-2.38$ to $z=-1.42$ Mm, while Figure \ref{f8}
shows similar cuts  for the layer $z=-1.42$ to  $z=-0.62$ Mm.

\begin{figure}
\epsscale{.8}
\plotone{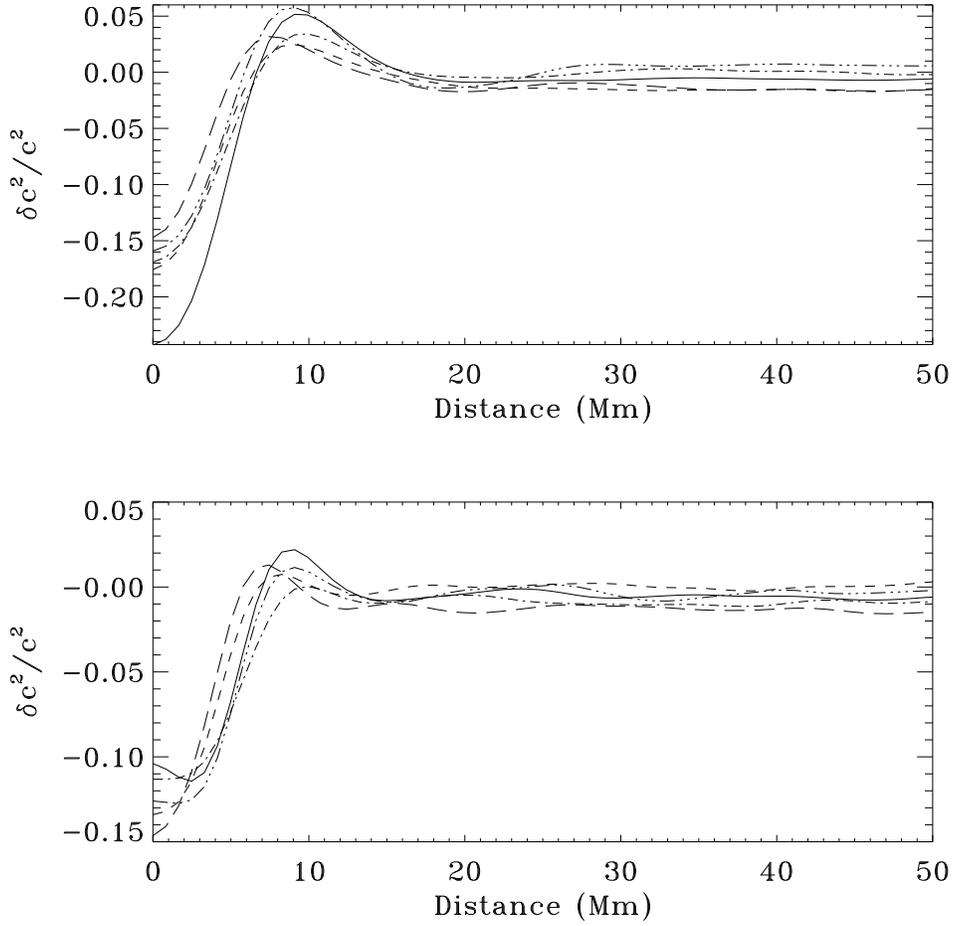}
\caption{Azimuthal averages of the inverted sound-speed perturbation at a depth of $z=-2.38$ to $z=-1.42$ Mm (upper panel) and $z=-1.42$ to $z=-0.62$ Mm (lower panel), as a function of the radial distance to the spot center, for 5 sunspots: ARs 9493 (solid line), 8073 (short-dashed line), 0615 (dot-dashed line), 8402 (dot-dot-dot-dashed line), and 0689 (long-dashed line).\label{f4}}
\end{figure}

An azimuthal  average of the  sound-speed perturbation for  5 circular
sunspots as  a function of the  radial distance to the  spot center is
plotted on Figure  \ref{f4}, for the layers $z=-2.38$  to $z=-1.42$ Mm
(upper  panel) and  $z=-1.42$ to  $z=-0.62$ Mm  (lower panel).  At the
layer  $z=-2.38$  to  $z=-1.42$  Mm,  the averaged  amplitude  of  the
increased sound-speed  region is always  above the noise level  on the
inverted results, with a  maximum value of about $\delta c^2/c^2=6\%$,
corresponding   to  a   sound-speed  increase   of  about   $0.45$  km
s$^{-1}$. Neglecting  the magnetic field influence, any  change in the
first adiabatic exponent, and any change in the mean molecular weight,
we have  $\delta c^2/c^2=\delta T/T$. Therefore, a $6\%$  increase in
$\delta c^2/c^2$,  translates into an upper limit  for the temperature
increase at $z=-1.9$ Mm of about  $1100$ K. For the layer $z=-1.42$ to
$z=-0.62$ Mm,  the maximum value  of the sound-speed  perturbation for
the 5 sunspots of Figure  \ref{f4} is about $\delta c^2/c^2=2\%$ after
azimuthal  averaging,   corresponding  to  an  upper   limit  for  the
temperature increase of $295$ K.

It is  worth mentioning that these  ring-like/arm-like structures have
already been detected  in the past: for instance,  Figure 4 of Hughes,
Rajaguru,  \& Thompson  (2005) clearly  shows a  ring on  the inverted
sound-speed  perturbation  map  at  $z=-2.3$  to  $z=-1.7$  Mm.  Their
inversion was performed using  ray-path approximation kernels, and the
algorithm  was LSQR  and not  the MCD  used here.  Another  example is
Figure 8  of Couvidat  et \mbox{al.} (2004)  which shows a  clear ring
structure  around   the  sunspot  NOAA  8243,   after  inversion  with
Fresnel-zone sensitivity kernels. The significance of these rings had been 
dismissed at that time.

\begin{figure}
\epsscale{.8}
\plotone{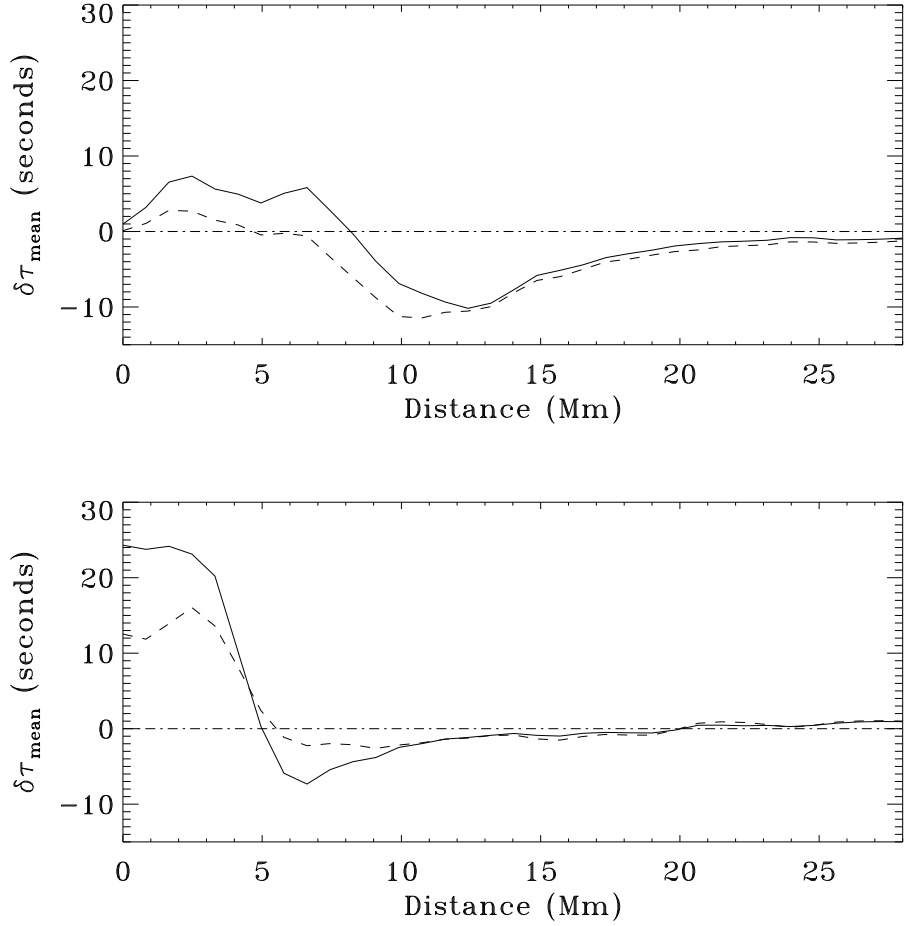}
\caption{Upper panel: azimuthal average of $\delta \tau_{\mathrm{mean}}(\mathbf{r},\Delta)$ for the distance $\Delta=11.6$ Mm and the active Region NOAA 8397. The solid line is before the power correction of Rajaguru et \mbox{al.} (2006), the dashed line is after correction. Lower panel: same plot for the active region NOAA 8073.\label{f11}}
\end{figure}

\begin{figure}
\epsscale{1.0}
\plotone{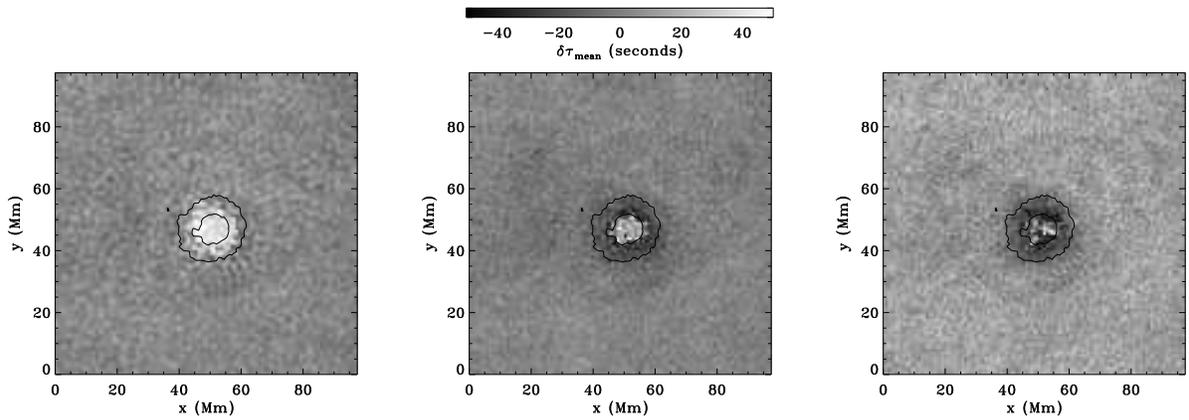}
\caption{Travel-time maps $\delta \tau_{\mathrm{mean}}(\mathbf{r},\Delta)$ for the distance $\Delta=11.6$ Mm and the active Region NOAA 9493. Left panel: with a frequency filter centered on 3 mHz; central panel: filter centered on 4 mHz; right panel: filter centered on $4.5$ mHz. The greyscale is the same for all three panels.\label{f10}}
\end{figure}

\begin{figure}
\epsscale{.8}
\plotone{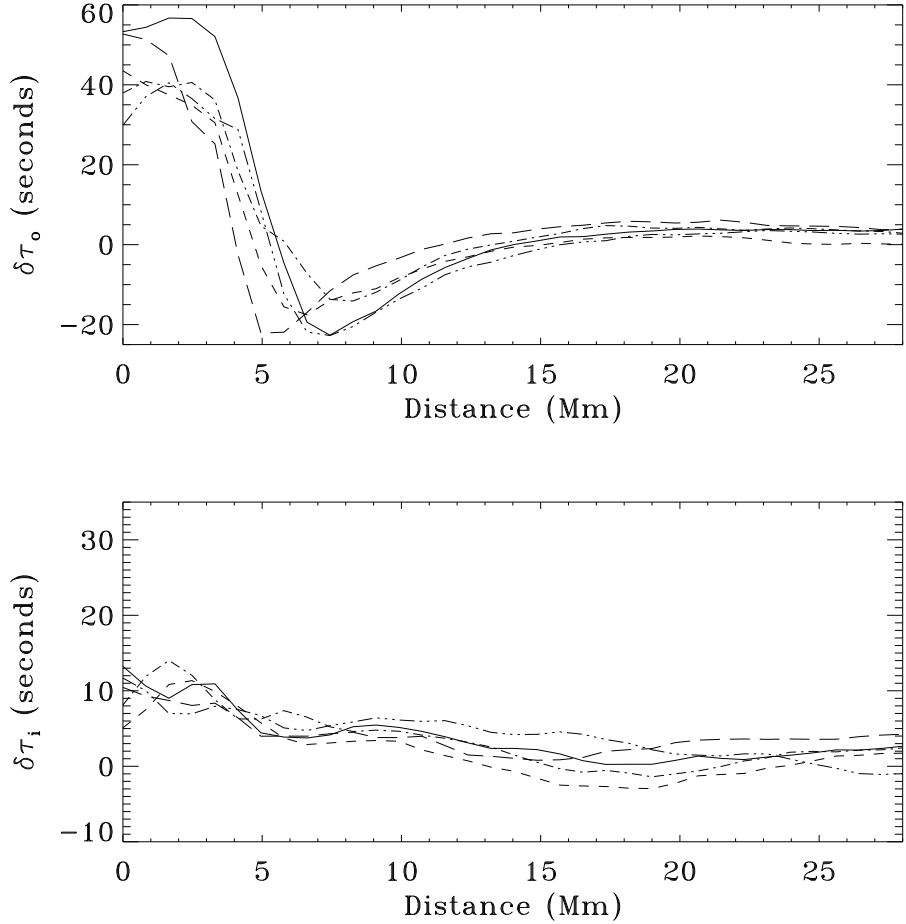}
\caption{Azimuthal average of $\delta \tau_{\mathrm{o}}(\mathbf{r},\Delta)$ (outgoing travel times; upper panel) and $\delta \tau_{\mathrm{i}}(\mathbf{r},\Delta)$ (ingoing travel times; lower panel) for $\Delta=11.6$ Mm and for 5 sunspots: NOAA 9493 (solid line), 8073 (short-dashed line), 0615 (dot-dashed line), 8402 (dot-dot-dot-dashed line), and 0689 (long-dashed line).\label{f13}}
\end{figure}

\begin{figure}
\epsscale{1.0}
\plotone{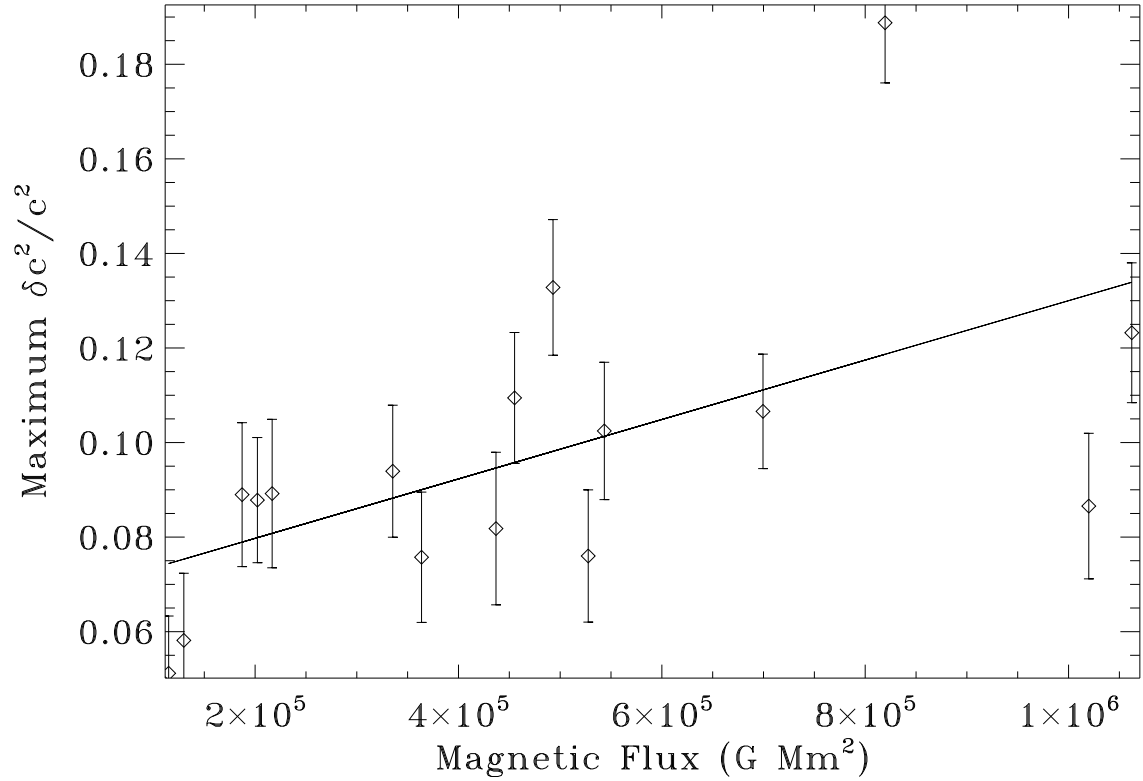}
\caption{Maximum value of the sound-speed perturbations in the rings surrounding the spots for the layer $z=-2.38$ to $z=-1.42$ Mm, as a function of the sunspot magnetic flux (see Table 1). The error bars are $\pm$ the rms variation of the inverted sound-speed perturbations at the selected layer. The solid line is the result of a linear fit.\label{f12}}
\end{figure}

To understand the nature of these rings ---artifacts or real physical features---
we perform several analyses. First, a recent study by Rajaguru  et \mbox{al.} (2006) shows that  the interaction of
phase-speed  filtering in  time-distance  analysis and  the spatially 
localized p-mode amplitude reduction in sunspots
introduces travel-time artifacts that are significant for short travel distances. 
To  make sure that the rings we observe
on the travel-time maps are not due to this p-mode amplitude effect, we applied
the power correction suggested  in their paper to a  couple of sunspots: the
dark ring is still present after correction, even though with a different size and a
different amplitude. Figure \ref{f11} shows the impact of this correction:
there might be a dependence on the sunpsot size; the dark ring being more 
affected by the power correction for small spots. 
Second, Braun  \& Birch (2006) observed frequency variations of solar p-mode 
travel times, using acoustic holography. They believe these variations are,
 at least partly, produced by surface effects due to magnetic fields.
To test whether or not the  dark rings have this frequency dependence, we
followed Braun  \& Birch (2006) and computed some  travel-time maps at  $\Delta=11.6$ Mm 
with the same phase-speed filter but different temporal frequency filters: 
on top of the phase-speed filter $F(k,\omega;\Delta)$ we add a Gaussian $\nu=\omega/(2\pi)$ 
filter with a FWHM of $\delta \nu=1$ mHz and centered on different frequencies ($\nu=$3, 4, and
$4.5$ mHz). The resulting travel-time maps for NOAA 9493 are shown in Figure \ref{f10}: 
the size and amplitude of the ring clearly depend on  the 
$\nu$ filter used. At low frequencies ($\nu$ filter centered on 3 mHz) the ring
is almost non-existent, but as the frequency content of the wave-packets shifts to higher
frequencies the ring broadens and its amplitude increases. 
At $4.5$ mHz, the ring extends inside the umbra and, maybe, even 
beyond the penumbra. Therefore, the near absence of rings on travel-time maps for 
$\Delta=6.2$ Mm might be partly due to the phase-speed filter we use at this distance: 
this filter selects wavepackets with a maximum power at about $\nu=3.5$ mHz. 
However, for $\Delta=11.6$ Mm, the selected wavepackets have a maximum power at about $4.25$ mHz.
It should be noticed that with our frequency and broad phase-speed filters,
selecting a specific frequency band also means selecting a slightly different average phase speed: both
effects are entangled. Therefore what we consider to be a dependence on frequency might partly
be a dependence on phase speed. Figure \ref{f10} also shows that the travel times in the umbra of the sunspot depend on frequency, 
albeit to a lesser extent than in the penumbra. Therefore, the rings might be the manifestation 
of a phenomenon affecting the entire spot, and directed from the periphery (the penumbra) to 
the center (the umbra) when the wave-packet frequency selected for the time-distance 
analysis is increased. Nevertheless, we do believe that 
the qualitative sound-speed inversion result below the umbra (two-region structure) holds, because 
the mean travel-time perturbation in the vicinity of the spot center is still consistently positive 
with the frequency filters applied for $\Delta=11.6$ Mm. However, this analysis undoubtedly confirms that the layers below 
a spot umbra have a lateral extent and an inverted $\delta c^2/c^2$ amplitude that depend on the 
wave-packet frequency. The problem seems less acute for deeper layers: the travel-time maps for 
larger distances $\Delta$ also vary with the frequency but no structure similar to the rings
is ever present. Even though we suggest that the qualitative result of the sound-speed inversion 
below a sunspot is valid, any attempt to quantitatively 
assess the corresponding sound-speed perturbation amplitude and the lateral extent of the decreased sound-speed region 
are inextricably connected to the frequency band selected. 
Third, with the phase-speed filters (no frequency filter), we notice a strong asymmetry between ingoing and outgoing 
travel-time maps: the rings seem to be present only on the outgoing travel times (see Figure \ref{f13}). 
This asymmetry might be an additional sign of an interaction of the wavefield with surface magnetic 
field (as suggested in, \mbox{e.g}, Lindsey \& Braun 2005), but such a feature can also be
 explained by a flow (as first suggested by Duvall et \mbox{al.} 1996). Figure \ref{f13} shows that the ring appears
both as partly due to a sound-speed perturbation (because it is visible on the mean travel-time maps),
and as partly due to a flow if we accept the interpretation of Duvall et \mbox{al.} (1996) (because it is visible 
on the difference travel-time maps). The sign of $\delta \tau_{\mathrm{o}}-\delta \tau_{\mathrm{i}}$
at $\Delta=11.6$ Mm points to an inflow below the spot umbra, and an outflow in the penumbral ring. Such a reversal in the 
flow direction seems physically implausible. Finally, Figure \ref{f12}  shows   a  loose  dependence  of   the  maximum  sound-speed
perturbation in these rings at a depth of $z=-2.38$ to $z=-1.42$ Mm as
a function of  the total magnetic flux of the  sunspot: the larger the
magnetic  flux or  the sunspot  size (Table  \ref{table.spot}  shows a
linear relation  between these two parameters), the  larger is $\delta
c^2/c^2$ in the arms/rings. This last observation is inconclusive though because
the subsurface structure of the spot might also depend on the total flux: it does not help
in discriminating between the ring-as-artifact and the ring-as-real-structure hypotheses.

Considering that the rings are located mainly below  the sunspot penumbra in the  inversion results and
inside  the penumbra  on the  travel-time maps, that there is a strong asymmetry 
between ingoing and outgoing travel-time maps (with rings present only on the outgoing travel times), 
and that the ring size and amplitude have a very clear frequency dependence, it seems hard to
believe that the sound-speed increase they imply is the signature of a
real solar  feature located in depth. Indeed  it seems more likely that
these  rings are spurious features  produced by frequency dependent surface effects
somewhat related to the strong magnetic field of the sunspots, which are not
accounted for in the inversion procedure.  

\section{CONCLUSION}

We clearly detect rings  of negative mean travel-time perturbations on
the  travel-time  maps of  most  sunspots  for  travel distances  between
$\Delta=8.7$  and $\Delta=11.6$  Mm. These  rings  produce significant
arm-like/ring-like  structures  on  the  inversion  results,  mimicking
regions  of increased sound speed.   However due  to the  locations of
these structures (immediately below  the sunspot penumbrae), and their
sensitivity  to   the  filtering  applied  to   the  Doppler  velocity
datacubes, it is likely  that they are artifacts produced by
the interaction  of the wavefield  and surface magnetic  fields. These
significant  structures were  not mentioned  in the  paper by  Zhao \&
Kosovichev (2006). Even though we agree with their conclusion that the
effect  of the  surface magnetic  field probably  does not  change the
qualitative inversion result of the basic sunspot structure (the two-region
structure), we show that these inversion results for the first 3-4
Mm below the solar surface  appear to be significantly contaminated by surface
effects probably of magnetic origin. Indeed, the rings could be the most visible
manifestation of a general phenomenon that also affects the amplitude and the lateral extent 
of the sound-speed perturbation in the shallower subsurface layers.

\section*{Acknowledgments}

This work  was supported by  NASA grants NNG05GH14G (MDI) \& NNG05GM85G (LWS). We
are grateful to A. C. Birch for providing us with his Born kernels, to
R. Wachter for fruitful discussions and  for teaching S. C. how to use
the fastrack software, to T. L. Duvall for his very useful advice, and to the anonymous
referee for improving this paper.

\clearpage

\begin{deluxetable}{ccccccccc}
\tablecaption{Characteristics of the Sunspots Studied, for a Threshold of 150 G \tablenotemark{a}\label{table.spot}}
\tablewidth{0pt}
\tablehead{\colhead{NOAA} & \colhead{Date} & \colhead{Area} & \colhead{Flux} & \colhead{Class} & \colhead{lat.} & \colhead{long.} & \colhead{Index}} 
\startdata
8073 & 08.17.1997 & 229.25    & 129897   & $\alpha$ & 14.5 & 282.0 & 40553\\
8243 & 06.18.1998 & 694.59    & 455180   & $\beta$  & 18.3 & 214.5 & 47871\\
8397 & 12.04.1998 & 745.76    & 492968   & $\beta$  & 16.5 & 140.6 & 51932\\
8402 & 12.05.1998 & 298.17    & 187218   & $\beta$  & 17.0 & 122.5 & 51958\\
8402 & 12.07.1998 & 788.06    & 436916   & $\beta$  & 16.8 & 121.8 & 51983\\
8602 & 06.29.1999 & 867.21    & 543457   & $\beta$  & 19.1 & 298.7 & 56893\\ 
8742 & 10.30.1999 & 796.93    & 527631   & $\beta$  &  7.0 & 121.0 & 59829\\
8760 & 11.10.1999 & 1677.79   & 1019900  & $\beta$  & 13.0 & 330.0 & 60102\\
9236 & 11.24.2000 & 1849.05   & 1062150  & $\beta$  & 18.5 & 359.0 & 69216\\
9493 & 06.12.2001 & 311.13    & 202267   & $\beta$  & 5.8  & 235.2 & 74032\\
0061 & 08.09.2002 & 576.55    & 363652   & $\beta - \delta$ & 8.5 & 45.2 & 84183\\
0330 & 04.09.2003 & 1288.19   & 819593   & $\beta - \gamma$ & 6.2 & 82.2  & 90015\\
0387 & 06.23.2003 & 617.49    & 335377   & $\beta - \gamma$ & 17.0 & 166.5 & 91806\\
0615 & 05.21.2004 & 404.61    & 216729   & $\alpha$ & 15.6 & 93.7 & 99812\\
0689 & 10.27.2004 & 206.06    & 114991   & $\alpha$ & 9.7 & 155.0 & 103621\\
0898 & 07.03.2006 & 1194.03   & 699494   & $\beta$  &-5.5 & 329.0 & 118363\\
\enddata
\tablenotetext{a}{NOAA is the active region number hosting the sunspot, as provided by the NOAA, the Area is in Mm$^2$, Flux is the l.o.s magnetic field flux in G Mm$^2$, \mbox{lat.} is the Carrington latitude and \mbox{long.} is the Carrington longitude of the center of the datacube where the active region is located, and Index is the MDI index of the first frame of the datacube retrieved on the SOI/MDI website. NOAA number and magnetic classes come from the website: http://www.solar.ifa.hawaii.edu/ARMaps/}
\end{deluxetable}

\clearpage

\clearpage

\clearpage

\clearpage

\clearpage

\clearpage

\clearpage

\clearpage

\clearpage

\clearpage

\clearpage

\end{document}